\documentclass[aps, pra, a4paper, amssymb, reprint, showpacs, nofootinbib,superscriptaddress]{revtex4-1}

\usepackage{amssymb}
\usepackage{graphicx}
\usepackage{amsmath}
\usepackage{hyperref}

\usepackage{bbold}
\usepackage{bbm}
\usepackage{color}

\newcommand{\abs}[1]{\left|#1\right|}
\newcommand{\sabs}[1]{|#1|}

\newcommand{\mD}{\mathcal{D}}
\newcommand{\mE}{\mathcal{E}}
\newcommand{\mA}{\mathcal{A}}

\newcommand{\mF}{\mathcal{F}}

\newcommand{\rmL}{\textrm{L}}

\newcommand{\figref}[1]{Fig.\,\ref{#1}}
\newcommand{\eeqref}[1]{Eq.\,\eqref{#1}}
\newcommand{\eqsref}[1]{Eqs.\,\eqref{#1}}
\newcommand{\secref}[1]{Sec.\,\ref{#1}}
\newcommand{\appref}[1]{Appendix \ref{#1}}

\newcommand{\vecb}[1]{\mathbf{#1}}

\newcommand{\rmd}{\textrm{d}}

\newcommand{\hc}{\textrm{h.c.}}
\newcommand{\rmat}{\textrm{at}}
\newcommand{\rmm}{\textrm{m}}
\newcommand{\rmf}{\textrm{f}}

\newcommand{\om}{\omega}
\newcommand{\Om}{\Omega}

\begin{document}
\date{\today}

\title{Long Distance Coupling of a Quantum Mechanical Oscillator to the Internal States of an Atomic Ensemble}

\author{B. Vogell}
\affiliation{Institute for Quantum Optics and Quantum Information, 
Austrian Academy of Sciences, Technikerstrasse 21a, 6020 Innsbruck, Austria}
\affiliation{Institute for Theoretical Physics, University of Innsbruck, Technikerstrasse 25, 6020 Innsbruck, Austria}

\author{T. Kampschulte}
\affiliation{Department of Physics, University of Basel, Klingelbergstrasse 82, 4056 Basel, Switzerland}

\author{M. T. Rakher}
\affiliation{Department of Physics, University of Basel, Klingelbergstrasse 82, 4056 Basel, Switzerland}

\author{A. Faber}
\affiliation{Department of Physics, University of Basel, Klingelbergstrasse 82, 4056 Basel, Switzerland}

\author{P. Treutlein}
\affiliation{Department of Physics, University of Basel, Klingelbergstrasse 82, 4056 Basel, Switzerland}

\author{K. Hammerer}
\affiliation{Institute for Theoretical Physics and Institute for Gravitational Physics, Leibniz University Hannover, Callinstrasse 38, 30167 Hannover, Germany}

\author{P. Zoller}
\affiliation{Institute for Quantum Optics and Quantum Information, 
Austrian Academy of Sciences, Technikerstrasse 21a, 6020 Innsbruck, Austria}
\affiliation{Institute for Theoretical Physics, University of Innsbruck, Technikerstrasse 25, 6020 Innsbruck, Austria}

\begin{abstract}
We propose and investigate a hybrid optomechanical system consisting of a micro-mechanical oscillator coupled to the internal states of a distant ensemble of atoms. The interaction between the systems is mediated by a light field which allows to couple the two systems in a modular way over long distances. Coupling to internal degrees of freedom of atoms opens up the possibility to employ high-frequency mechanical resonators in the MHz to GHz regime, such as optomechanical crystal structures, and to benefit from the rich toolbox of quantum control over internal atomic states. Previous schemes involving atomic motional states are rather limited in both of these aspects. We derive a full quantum model for the effective coupling including the main sources of decoherence. As an application we show that sympathetic ground-state cooling and strong coupling between the two systems is possible.
\end{abstract}


\maketitle

\section{Introduction}
In this paper we describe an optical interface which provides a coherent quantum mechanical coupling between a nano-mechanical oscillator and the internal states of an atomic ensemble. The motivation for considering such a hybrid quantum device should be seen in the context of the effort to build composite quantum systems, where complementary advantages of the components are combined in a single, experimentally compatible setup. In recent years various hybrid systems involving nano-mechanical oscillators have been investigated, including mechanical oscillators coupled to solid-state spin systems \cite{Degen:2009kd,Arcizet:2011cg,Kolkowitz:2012iw,Teissier:2014gt,Ovartchaiyapong:2014gv}, semiconductor quantum dots \cite{Yeo:2013ja,Montinaro:2014im}, superconducting devices \cite{Oconnell:2010br,Pirkkalainen:2013gh}, as well as cold atoms \cite{Wang:2006dd,Hunger:2010fr,Camerer2011,Jockel:2014kx}. In the context of nano-mechanics such hybrid devices provide novel opportunities for cooling, detection and quantum control of vibrations in engineered mechanical structures, with applications in precision sensing and fundamental tests of quantum physics \cite{Treutlein:2012wa,Hunger:2011eo,Aspelmeyer:2012fy,Aspelmeyer:2013vr}.  The nano-mechanics -- atomic ensemble hybrid system developed in the present work takes advantage of the well-developed atomic toolbox to manipulate atomic systems with lasers  \cite{Weidemueller:2009}. At the same time light as the mediator of interactions provides the unique opportunity for coupling {\em distant} quantum systems, in the present example a nano-mechanical oscillator in a cryogenic environment and an atomic ensemble in an cold atom chamber.

Previous work on coupling nano-mechanical oscillators to atoms has focused mainly on coupling to the {\em motional} degrees of freedom of the atoms, where the atoms act as a microscopic mechanical oscillator deep in the quantum regime. In this context, various coupling mechanisms have been proposed \cite{Hammerer2010,Hunger:2011eo,Vogell2013,Bennett:2014ui}, and recently first experimental implementations have been reported \cite{Wang:2006dd,Hunger:2010fr,Camerer2011,Jockel:2014kx}. In particular, substantial sympathetic cooling of a mechanical oscillator by coupling it to the laser-cooled motion of an ensemble of ultracold atoms has been observed \cite{Jockel:2014kx} with an exciting prospect to achieve ground-state cooling \cite{Vogell2013}. In the quest to establish more advanced levels of quantum control in such a hybrid system the coupling to the center of mass motion of atoms is challenged through two limitations: Firstly, the requirement of resonant coupling limits the frequency of the mechanical oscillator to the maximal trap frequency achievable in optical lattices, that is, to the sub-MHz regime. Secondly, while motional states of individual atoms in optical lattices are under complete control, a similar level of quantum control over the center of mass motion of atomic ensembles has yet to be established. 

In the present work we will consider the coupling of a nano-mechanical oscillator to the {\em internal} states of the atomic ensemble. Coupling to Zeeman or hyperfine ground states with frequencies in the MHz up to GHz regime opens up the possibility to use high-frequency mechanical oscillators, such as optomechanical crystal structures \cite{Cohen13,Eichenfield09}, which generically exhibit much larger radiation pressure coupling to light. Coupling to internal degrees of freedom also benefits from the rich toolbox available for the manipulation, initialization and measurement of the electronic atomic states with laser light. Moreover, internal states of atomic ensembles can realize effective mechanical oscillators with unusual properties such as negative mass \cite{Polzik:2014vp} or reduced quantum uncertainty through spin-squeezing \cite{Ockeloen:2013wva}. Alternatively, manipulation and state-selective detection on the level of single quanta is possible using techniques of Rydberg blockade \cite{Carmele:2014hl}.

Here we show that long-distance coupling of a mechanical oscillator to the internal states of an atomic ensemble is possible. Related work suggesting long-distance coupling to internal levels has been reported in \cite{Hammerer2009a,Bariani:2014tr}. We derive a full quantum mechanical theory for a specific experimentally relevant geometry, including the derivation of the coherent coupling, the discussion of quantum noise and the complete dynamics resulting from the quantum stochastic Schr{\"o}dinger treatment. A quantized many-body treatment is essential as the coupling and dissipation channels may be modified by collective effects that cannot be obtained in semi-classical or single-particle theories. The dynamics of our coupled mechanical-atomic ensemble system are exactly solvable, which allows an estimate of a parameter regime for sympathetic cooling and strong coupling. Considering a photonic crystal ``zipper cavity'' as the optomechanical device \cite{Cohen13,Eichenfield09}, we obtain significantly faster dynamics and better performance than in previous motional-state coupling schemes \cite{Vogell2013}. The paper is structured as follows: In \secref{sec:model}  we present the full quantum model of the light-mediated coupling and the main decoherence processes. In \secref{sec:applications} we propose different applications such as sympathetic cooling of the mechanical oscillator and strong atom-oscillator coupling, while experimental parameters are discussed in \secref{sec:parameters}.

\section{Model}
\label{sec:model}
\begin{figure}[h!]
\centering
\includegraphics[width=0.5\textwidth]{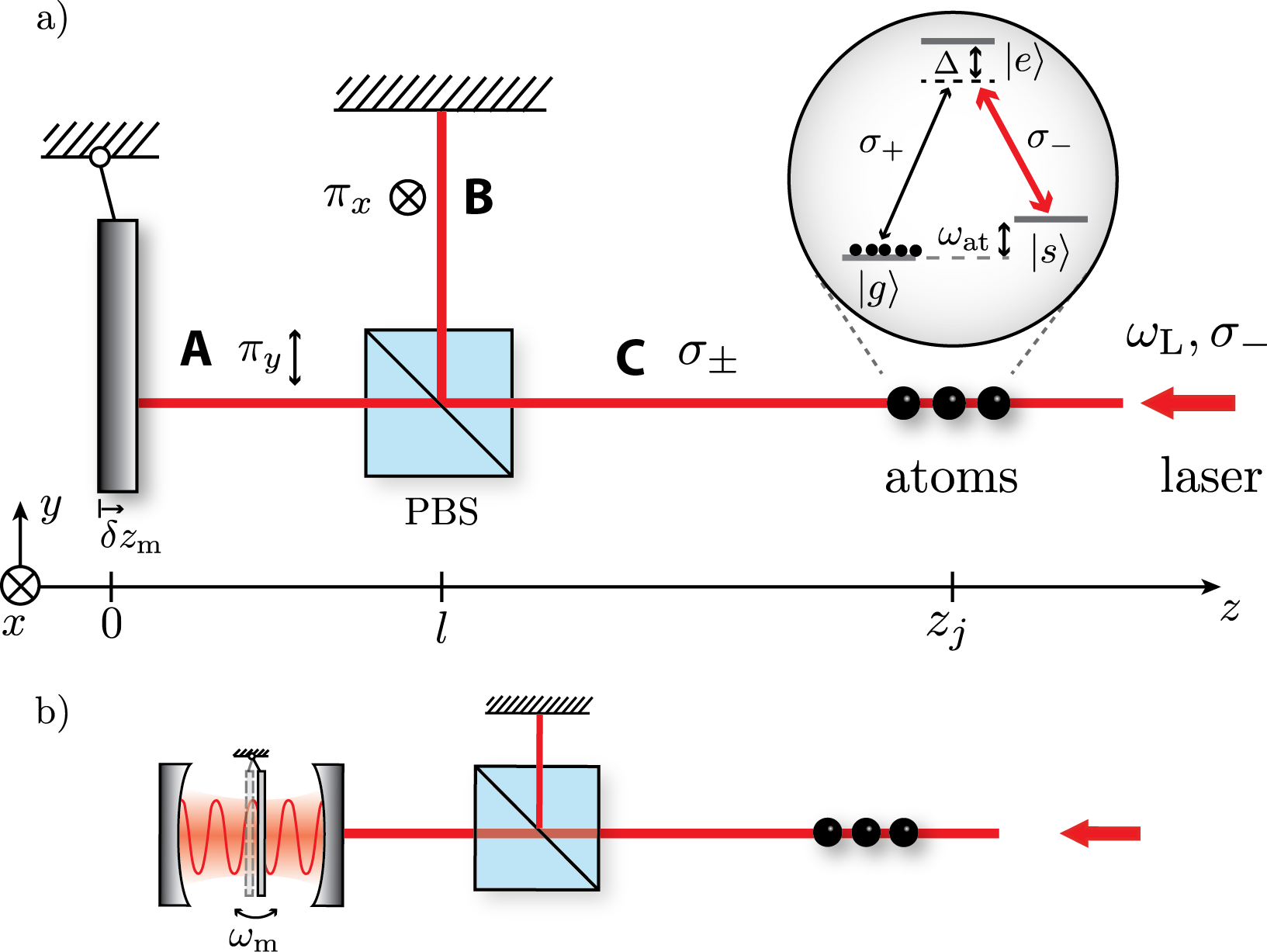}
\caption{(a) Micro-mechanical mirror coupled to the internal states of a distant atomic ensemble mediated by the light field (red). The system is pumped by a $\sigma_-$-polarized laser at frequency $\omega_\rmL$ and amplitude $\alpha$. 
The Michelson-interferometer like setup with the PBS is needed to translate the motion of the mechanical resonator into a polarization rotation at the position of the atoms.  
In the other direction, transitions between the groundstates $(|s\rangle, |g\rangle)$  change the radiation pressure on the movable mirror. 
{\it Inset}: atomic level scheme: $|s\rangle$ and $|g\rangle$ are separated by $\omega_{\rm at}$. All atoms are initially prepared in $|g\rangle$, while the incoming laser coherently pumps the $|s\rangle \leftrightarrow |e\rangle$ transition with a detuning $\Delta$. 
(b) Extension of the setup: "membrane-in-the-middle``-setup \cite{Thompson:2008dx}, where the mechanical resonator is placed inside a cavity to increase the coupling by its finesse $\mathcal{F}$. Similarly, any other optomechanical system with a single sided cavity can be implemented.
}
\label{fig:fig1}
\end{figure}

We consider a system as shown in \figref{fig:fig1}(a), where a micro-mechanical resonator (left) is coupled via the light field to the internal states of a distant atomic ensemble. 
The atomic ensemble is trapped in an external optical lattice and consists of $N$ three-level atoms with a $\Lambda$-type level scheme as depicted in the inset of \figref{fig:fig1}(a), where the two ground states ($|g\rangle$, $|s\rangle$) are separated by $\omega_{\rm at}$, and the corresponding transitions to the excited state $|e\rangle$ are polarization-dependent. Initially, the dominant population of the atoms is prepared in state $|g\rangle$.
At the position of the atoms the light field is, on average $\sigma_{-}$-polarized, since it is pumped by a $\sigma_-$-polarized laser at frequency $\omega_\rmL$ from the right with amplitude $\alpha$. The latter is related to the running wave power $P=\hbar \om_{\rm L} \alpha^2/2\pi$ of the laser, which drives the transition $|s\rangle \leftrightarrow |e\rangle$ off-resonantly with detuning $\Delta=\omega_\rmL-\omega_{\rm es}$. 

In a Michelson interferometer-like setup, a polarizing beamsplitter (PBS) splits the circularly polarized light into linearly polarized light on arm $A$ ($\pi_{y}$) and $B$ ($\pi_{x}$). 
In arm $A$, the mechanical resonator is taken to be a perfect mirror with effective mass $M$ and resonance at frequency $\omega_\rmm$, such that its zero-point fluctuations are given by $\ell_\rmm=\sqrt{\hbar/2M\omega_\rmm}$. 
The second arm of the Michelson interferometer-like setup, arm $B$ in \figref{fig:fig1}(a), is bounded by a fixed mirror at position $x=l$ and has equal length to arm $A$ as long as the mechanical resonator is in its equilibrium position. This also ensures that the outgoing light has predominantly the same polarization as the incoming light.
Note, while we describe in the following the simple minded setup of a moving end-mirror as depicted in \figref{fig:fig1}(a), it is straight forward to use more sophisticated setups like a "membrane-in-the-middle`` configuration as displayed in \figref{fig:fig1}
(b), or even a fully one dimensional setup without the Michelson interferometer.

The coupling of the mechanical resonator to the atoms works via translating the phase shift caused by a displacement of the mechanical resonator into a polarization rotation using the polarizing beamsplitter. 
In case of resonance $\omega_{\rm at} \approx \omega_{\rm m}$, the emergent $\sigma_+$-polarized light on arm $C$ at the blue sideband frequency $\omega_{+}=\om_{\rm L}+\om_{\rm m}$ can then induce a two-photon transition on the side of the atoms, i.e. $|s\rangle \leftrightarrow |g\rangle$. 
In return, if the atoms make a transition between the two ground states, the radiation pressure on the mirror changes due to the additional emitted photons of $\sigma_+$-polarization which have $50\%$ chance to enter arm $A$.

In the following we derive a quantum-mechanical description for the coupling of the mechanical resonator to the internal states of the atomic ensemble.

\subsection{Mode Functions}
\label{sec:modefunctions}

We start the quantum mechanical treatment by quantizing the field modes for the case where the mirror is in its equilibrium position, such that the two arms $A$ and $B$ of the Michelson interferometer-like setup have equal length. 
There are two sets of field modes representing the two possible polarizations incident from the right on arm $C$. We choose the basis of circular $\sigma_{\pm}$-polarized light with the associated destruction operators being $c_\omega (\sigma_-)$ and $d_\omega (\sigma_+)$, that obey the commutation relations $[c_\omega,c_{\omega'}^\dag]=[d_\omega,d_{\omega'}^\dag]=\delta(\omega-\omega')$.
The PBS then decomposes these circularly polarized modes into linearly polarized ones on arm $A$ ($\pi_{y}$) and $B$ ($\pi_{x}$). 

Taking into account the boundary conditions at both mirrors, the positive frequency parts of the electric field are given by
\begin{align}
\textrm{on A:}
\quad &\vecb{E}^{(+)}(z)=\vecb{e}_y\int\rmd\om\,\mE_\om \frac{i}{\sqrt{2}}(c_\om+d_\om)\sin(kz) \label{eq:modesA}\\
\textrm{on B:}
\quad &\vecb{E}^{(+)}(y)=\vecb{e}_x\int\rmd\om\,\mE_\om \frac{1}{\sqrt{2}}(c_\om-d_\om)\sin(ky)\\
\textrm{on C:}
\quad &\vecb{E}^{(+)}(z)=
\vecb{e}_{+} E_{\sigma_+}^{(+)}(z)+\vecb{e}_{-} E_{\sigma_-}^{(+)}(z)\\
&E_{\sigma_-}^{(+)}(z)
= \int\rmd\om\, \mE_\om c_\omega \sin(kz)\label{eq:sigmaplusC}\\
&E_{\sigma_+}^{(+)}(z)
= \int\rmd\om\, \mE_\om d_\omega \sin(kz)\label{eq:sigmaminusC},
\end{align}
where $k=\omega/c$, $\mE_\om=\sqrt{\hbar\om /\pi c \epsilon_0 \mA}$. The beam cross-sectional area $\mA$ is in principle a function of the position and can therefore be different at the position of the atoms and the mechanical oscillator. We will use the same letter $\mA$ for the two cases as it is clear from the context what we refer to.
Further, $\vecb{e}_{i}$ with $i\in \{x,y\}$ are the polarization unit vectors for linear polarized light,
 and $\vecb{e}_{\pm}=\mp (\vecb{e}_x \pm i \vecb{e}_y)/\sqrt{2}$ the associated ones for circular polarized light.

\subsection{Hamiltonian}
\label{sec:Hamiltonian}

The full system in a one-dimensional (1D) model is described by the Hamiltonian  
\begin{align}
H   &= H_0 + H_{\rm m-f} + H_{\rm at-f},
\end{align}
where $H_{\rm m-f} $ describes the interaction between the light field and the mechanical oscillator and $H_{\rm at-f}$ is the interaction of the atomic ensemble with the light field.
$H_0$ contains the free evolution of the mechanics, the energy of the atomic ground states and the field modes:
\begin{align}
\label{eq:H0}
H_0 &=\hbar \om_\rmm a_\rmm^\dag a_\rmm +\hbar \omega_{\rm at} \sum_j \sigma^j_{\rm ss}+H_{\rm field},
\end{align}
where the field mode Hamiltonian reads $H_{\rm field}= \hbar \int\! \rmd\om\,\om(c_\om^\dag c_\om + d_\om^\dag d_\om)$. Further, we set the energy of ground state $|g\rangle$ to zero and the splitting of the two ground states is given by $\omega_{\rm at}$, see inset of \figref{fig:fig1}(a). 

The interaction between the mirror and the light-field $H_{\rm m-f}$ is modeled by the familiar radiation pressure Hamiltonian, which can be derived from the Maxwell stress tensor. For reasons of simplicity, here we evaluate the Maxwell stress tensor for an ideal metallic mirror.
However, we note that a physical equivalent interaction can also be derived for other mechanical systems such as a "membrane-in-the-middle``-configuration as depicted in \figref{fig:fig1}(b), cf. \secref{sec:genericOMsystem} and \cite{Vogell2013}.
The interaction Hamiltonian for the ideal metallic mirror reads
\begin{align}
H_{\rm m-f}=\frac{\mA}{\mu_0} \vecb{B}^{(-)}_A(0) \vecb{B}^{(+)}_A(0) \delta z_\rmm, 
\label{eq:HRP}
\end{align}
where the displacement of the mirror is $\delta z_\rmm= \ell_\rmm (a_\rmm+a_\rmm^\dag)$ with mechanical annihilation (creation) operator $a_\rmm^{(\dag)}$. 
The positive frequency part of the magnetic field on arm $A$ is given by
\begin{align}
\vecb{B}^{(+)}_A(z)=\vecb{e}_x\int\rmd\om\,\frac{-\mE_\om}{c\sqrt{2}}(c_\om+d_\om)\cos(kz). 
\label{eq:QM_Bfield} 
\end{align}

On the side of the atoms we assume a level scheme as shown in the inset of \figref{fig:fig1}(a), where each polarization ($\sigma_\pm$) couples to one arm of the $\Lambda$-transition. 
Considering the Hamiltonian for the $\Lambda$-system interacting with the light field, we first eliminate the excited state $|e\rangle$ that is detuned by $\Delta$ with respect to the laser. Here, the condition $\Omega = \mu \alpha \mE_{\om_\rmL}/\hbar \ll \sabs{\Delta}$ enters, where $\alpha$ denotes the laser amplitude which will be introduced in \eeqref{eq:displace}.
Subsequently, we obtain the effective interaction of the two ground states ($|g\rangle$, $|s\rangle$) with the light field
\begin{align}
H_{\rm at-f}
=&\frac{\mu_+ \mu_-}{\hbar \Delta} \sum_j \left[
  E^{(+)}_{\sigma_-}(z_j) E^{(-)}_{\sigma_+}(z_j) \sigma_{\rm gs}^j    \right. \notag \\
  &\hspace{1.7cm} +  E^{(+)}_{\sigma_+}(z_j) E^{(-)}_{\sigma_-}(z_j)\sigma_{\rm sg}^j
 \Big] \notag \\
&+ \frac{\mu_+^2}{\hbar \Delta} \sum_j E^{(+)}_{\sigma_+}(z_j) E^{(-)}_{\sigma_+}(z_j)\sigma_{\rm gg}^j \label{eq:Hatf} 
 \\
& +\frac{\mu_-^2}{\hbar \Delta} \sum_j E^{(+)}_{\sigma_-}(z_j) E^{(-)}_{\sigma_-}(z_j)\sigma_{\rm ss}^j,   \notag
 \end{align}
where $\mu_\pm$ are the atomic dipole matrix elements for both transitions, $z_j$ is the position of the $j$th atom, and $\sigma_{ab}=|a\rangle\langle b|$ is the atomic transition operator. 
In \eeqref{eq:Hatf} the first lines provide the relevant interaction, whereas the last two lines denote the ac-Stark shifts for both ground states.

\subsection{ Linearization around the laser} 
\label{sec:linearization}

We now introduce the $\sigma_-$-polarized laser displayed in \figref{fig:fig1}(a), which mediates the coupling between the mechanical resonator and the atoms. The $\sigma_-$-polarized light field then contains a coherent part plus fluctuations.
 To include this, we move to a displaced picture by applying the following replacement for the field modes,
\begin{align}
c_\omega\rightarrow c_\omega  +  \alpha \delta(\omega-\omega_L) e^{-i\om_L t}
\label{eq:displace}
\end{align}
with the amplitude $\alpha$ and laser frequency $\omega_L$. 
 Assuming $\sabs{\alpha}\gg1$ allows us to linearize the interactions $H_{\rm at-f}$ and $H_{\rm m-f}$ by only keeping contributions enhanced by $\alpha$.

\subsubsection{Mirror-field interaction} 

We start with the mirror-field interaction in \eeqref{eq:HRP} by inserting the magnetic field in \eeqref{eq:QM_Bfield}, and apply the above replacement to the associated field mode operators. 
The contribution $\propto \alpha^2$ is taken care of by redefining the equilibrium position of the mirror, since it yields only a constant force. 
The zeroth order in $\alpha$ is neglected, while the linear order provides the relevant interaction, which is in an interaction picture with respect to (w.r.t.) $H_{\rm field}$ 
\begin{align}
H_{\rm m-f}^{\rm lin}=&\hbar g_\rmm \left[ c(t) + d(t)+ c^\dag(t)+d^\dag(t) \right] X_\rmm,
\label{eq:Hlinmf}
\end{align}
where we defined the mechanical quadrature as $X_\rmm=(a_\rmm+a_\rmm^\dag)/\sqrt{2}$ and the mirror-light coupling element for $\mE_\omega\approx \mE_{\om_L}$ reads 
\begin{align}
g_\rmm
=\frac{\alpha k_L \ell_\rmm}{\sqrt{ \pi}}.
\label{eq:gm}
\end{align}
Further, we defined the field mode operators as
\begin{align}
c(t)&=\int_{\om_{\rm L}-\theta}^{\om_{\rm L} + \theta} \frac{\rmd\om}{\sqrt{2\pi}}\,c_\om\, e^{-i(\om-\om_{\rm L})t} \label{eq:Defc}\\
d(t)&=\int_{\om_{\rm L}-\theta}^{\om_{\rm L}+ \theta} \frac{\rmd\om}{\sqrt{2\pi}}\,d_\om\, e^{-i(\om-\om_{\rm L})t}\,.\label{eq:Defd}
\end{align}
The associated commutation relations are given by 
\begin{align}
\label{eq:commutationrel}
[c(t),c^\dag(t')]=[d(t),d^\dag(t')]&=\delta_{\theta}(t-t'),
\end{align}
where $\delta_{\theta}(t)$ is a representation of the $\delta$-function of width $\propto 1/\theta$. Here, we assume that all photons mediating the relevant interaction processes have a frequency in a bandwidth $2\theta$ around the laser frequency $\omega_{\rm L}$, such that the frequency scales satisfy the condition $\omega_\rmm, \omega_{\rm at} \ll \theta \ll \omega_{\rm L}$.

\subsubsection{Atom-field interaction} 
In order to linearize the atom-field interaction $H_{\rm at-f}$ we insert the electric field in \eqsref{eq:sigmaplusC}-\eqref{eq:sigmaminusC} into \eeqref{eq:Hatf} and apply the displacement of the $\sigma_{-}$-field mode given in \eeqref{eq:displace}. 
The classical contribution that is quadratic in $\alpha$ yields an optical lattice for atoms in the $|s\rangle$-state given by
\begin{align}
H_{\rm trap}=\hbar  \,\Omega_{\rm OL}  \sum_j \sin^2 (k_L z_j) \,\sigma^j_{\rm ss}
\label{eq:optLatsstate}
\end{align}
with frequency
\begin{align}
\Omega_{\rm OL}=\frac{\mu_-^2}{\hbar^2 \Delta} \alpha^2 \mE^2_{\omega_L}.
\end{align}
Since this optical lattice traps only the $|s\rangle$-state and the majority of population is in the $|g\rangle$-state, an external optical lattice that traps both states is necessary. In \secref{sec:positioning} we specify the conditions for the external optical lattice. \\

The relevant contribution linear in $\alpha$ provides us with the interaction between atoms and light field. Therefore, we first transform into an interaction picture w.r.t. $H_{\rm field}$ and approximate $\mE_{\om} \approx \mE_{\om_L}$ as well as $\tau\equiv\tau_j=z_j/c$, the retardation time due to the propagation of light between atoms and mechanical oscillator. The latter approximation means the retardations across the atomic ensemble are neglected in the following.
With this, we can write\footnote{Note, that linearizing the last line in \eeqref{eq:Hatf} also yields a contribution linear in $\alpha$ that is not resonant with the interaction. Further, this contribution couples to the population of the $|s\rangle$-state, which is weak in this case, and therefore we have neglected this contribution in \eeqref{eq:Hlinatfcontrib}.}
\begin{align}
H_{\rmat-\rmf}^{\rm lin}
 = \hbar g_\rmat \sum_j &\left[ \Big\{ d^\dag(t-\tau)\left(1-e^{-2i k_L z_j}\right)   \right.    \label{eq:Hlinatfcontrib}
 \\
 & \left. +d^\dag(t+\tau)\left(1-e^{2i k_L z_j}\right) \Big\} \sigma^j_{\rm gs}
 + \hc \right], \notag
 \end{align}
where we used \eeqref{eq:Defd} and introduced the atom-field coupling 
\begin{align}
g_{\rm at}= \frac{\mu_+ \mu_-}{\hbar^2 \Delta} \alpha \, \mE_{\omega_L}^2 \sqrt{\frac{\pi}{8}}.
\end{align}

\begin{figure}
\centering
\includegraphics[width=0.45\textwidth]{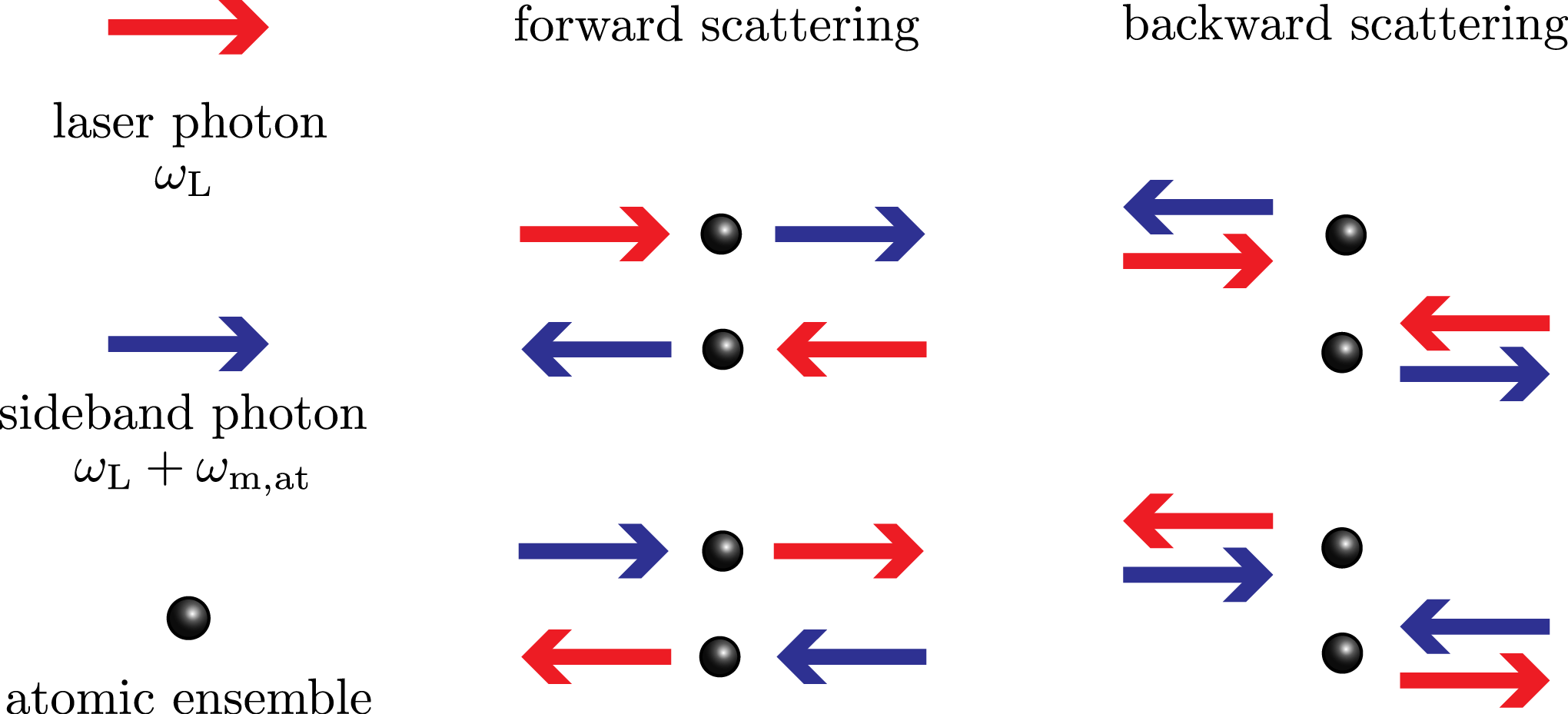}
\caption{ Interactions between the atomic ensemble and the light field are given by Stokes scattering processes, i.e. conversion between laser photons and sideband photons. Forward scattering processes (left column) are associated with the unmodulated spin wave $S_0$ in \eeqref{eq:AEbasis}. On the other side, backward scattering processes (right column) result in the two modulated spin waves $S_{\pm 1}$ in \eeqref{eq:AEbasis}.}
\label{fig:Hatf}
\end{figure}

Interpreting \eeqref{eq:Hlinatfcontrib} we see that every spin flip requires a two photon process of either absorbing a $\sigma_-$-polarized laser photon at frequency $\omega_{\rm L}$ and emitting a $\sigma_+$-polarized sideband photon at frequency $\omega_{\rm L}+\om_\rmat$, or vice versa. From \eeqref{eq:Hlinatfcontrib}  we see, that these two processes can either result in forwards scattering or in backwards scattering and can occur at different times $t\pm \tau$ as displayed in \figref{fig:Hatf}.
 
We are interested in a description of the interaction between the light field and the collective spin excitation of the atomic ensemble, i.e. a spin wave. 
In order to obtain such an interaction, we rewrite \eeqref{eq:Hlinatfcontrib} in the collective atomic excitation states given by 
\begin{align}
S_{0}=\frac{1}{\sqrt{N}} \sum_{j} \sigma_{\rm gs}^{j} \qquad S_{\pm 1}=\frac{1}{\sqrt{N}} \sum_{j} \sigma_{\rm gs}^{j} e^{\pm 2 i k_{\rm L}z_{j}}
\label{eq:AEbasis},
\end{align}
where $S_0$ corresponds to an unmodulated spin wave resulting from forwards scattering of the photon. Backwards scattering of the photon leads to a space-dependent phase and therefore results in modulated collective spin waves $S_{\pm 1}$. 
By using \eeqref{eq:AEbasis} the atom-field Hamiltonian reads 
\begin{align}
H^{\rm lin}_{\rm at-f}
=&\hbar \sqrt{N} g_\rmat \Big[ d^\dag(t-\tau) \left( S_0 -S_{-1} \right) \label{eq:Hatfcollbas1}\\
& \hspace{1.25cm}  + d^\dag(t+\tau) \left( S_0-S_{+1}\right)   + \hc
  \Big].  \notag  
\end{align}

Although the collective modes in \eeqref{eq:AEbasis} are intuitive as they correspond to forward and backward scattering, it is sufficient to describe the interaction with just two spin waves. Therefore, we introduce the collective basis
\begin{align}
S_{\rm sin}&=\frac{1}{\sqrt{N}} \sum_{j} \sin(2 k_{\rm L} z_j) \sigma_{\rm gs}^{j} \notag \\
S_{1-{\rm cos}}&=\frac{1}{\sqrt{N}} \sum_{j} \left\{1- \cos(2 k_{\rm L} z_j) \right\} \sigma_{\rm gs}^{j}.
\label{eq:AEbasisCS}
\end{align}
 \eeqref{eq:Hatfcollbas1} can then be expressed as
\begin{align}
H^{\rm lin}_{\rm at-f}
=&\hbar \sqrt{\frac{N}{2}} g_\rmat \Big[ X_{d}(t+\tau) \Big\{ S_{1-{\rm cos}}+S_{1-{\rm cos}}^\dag \notag  \\
&\hspace{3cm}+i( S_{{\rm sin}}^\dag-S_{{\rm sin}})\Big\} \notag  \\
&\hspace{1.2cm} +P_{d}(t+\tau) \Big\{ i( S_{1-{\rm cos}}^\dag-S_{1-{\rm cos}})  \notag  \\
&\hspace{3cm}-S_{{\rm sin}}-S_{{\rm sin}}^\dag\Big\} \notag  \\
&\hspace{1.2cm} +X_{d}(t-\tau) \Big\{ S_{1-{\rm cos}}+S_{1-{\rm cos}}^\dag  \notag  \\
&\hspace{3cm}-i( S_{{\rm sin}}^\dag-S_{{\rm sin}})\Big\} \notag  \\
& \hspace{1.2cm} +P_{d}(t-\tau) \Big\{ i( S_{1-{\rm cos}}^\dag-S_{1-{\rm cos}})   \notag  \\
&\hspace{3cm}+S_{{\rm sin}}+S_{{\rm sin}}^\dag\Big\} \Big],
\label{eq:Hatfcollbas3} 
\end{align}
where we introduced the light field quadratures 
\begin{align}
X_{d}(t)&= \left\{d(t)+d^\dag(t) \right\}/\sqrt{2} \notag \\
 P_{d}(t)&=i \left\{d^\dag(t)-d(t) \right\}/\sqrt{2}.
 \label{eq:XPdfield}
\end{align}
Note, \eqsref{eq:XPdfield}  represents canonical operators fulfilling the associated commutation relations, i.e $[X_d(t),P_d(t')]=i \delta(t-t')$.

\subsubsection{Positioning of the atoms} 
\label{sec:positioning}

So far no assumptions on the trapping of the atomic ensemble were made. 
In contrast to previous proposals \cite{Vogell2013}, where the coupling laser also provided the optical lattice that traps the atomic ensemble, here we have only a space-dependent Stark shift of the $|s\rangle$-state due to the coupling laser, cf. \eeqref{eq:optLatsstate}. 

In the following, we want to simplify the atom-light field interaction in \eeqref{eq:Hatfcollbas3}, which couples to the two spin waves of the atomic ensemble from \eeqref{eq:AEbasisCS}, such that both spin waves reduce to the same unmodulated spin wave $S_0$ in \eeqref{eq:AEbasis}.
In order to do so, 
we introduce an optical lattice that traps both ground states of the atomic ensemble. 
This can be obtained by choosing an appropriate optical lattice localizing the atoms at positions $z_j$ such that $\sin(2 k_L z_j)=1$. 

Since the Stark shift of the $|s\rangle$-state in \eeqref{eq:optLatsstate} has the same spatial dependence, it reduces to a constant shift that can easily be compensated and is equal for all atoms.

By introducing the positioning as above only the unmodulated spin wave $S_0$ is relevant, since both modulated spin waves $S_{\pm 1}$ reduce to $S_0$. 
The quadratures of the unmodulated spin wave are given by
\begin{align}
X_{s,0}&=\left(S_0+S_0^\dag\right)/\sqrt{2} \notag \\
P_{s,0}&=i \left(S^\dag_0-S_0\right)/\sqrt{2}.
 \label{eq:XPspinwave}
\end{align}
Under the conditions that the dominant population of the atoms is occupying the ground state $|g\rangle$ and by applying the Holstein-Primakoff approximation \cite{Holstein1940}, the quadratures in \eeqref{eq:XPspinwave} fulfil the canonical commutation relations $[X_{s,0},P_{s,0}]\approx i$.

\subsection{Linearized Hamiltonian} 
\label{sec:linHamiltonian}

Finally, we can summarize the resulting linearized Hamiltonian that contains both the coupling of the light field to the mechanical resonator and the atomic ensemble, respectively. 
Including the statements of positioning the atoms in \secref{sec:positioning}, the complete linearized Hamiltonian in an interaction picture w.r.t. $H_{\rm field}$ is given by
\begin{align}
H_{\rm lin}
=&H_{0,{\rm lin}}+ \hbar \sqrt{2} g_\rmm \left[ X_{c}(t)+X_{d}(t)\right] X_\rmm \label{eq:Hlin} \\
&+\hbar \sqrt{N} g_\rmat \Big[ X_{d}(t+\tau) \Big\{ X_{s,0} +P_{s,0}\Big\} \notag  \\
&\hspace{1.6cm} +P_{d}(t+\tau) \Big\{ P_{s,0} -X_{s,0}\Big\} \notag  \\
&\hspace{1.6cm} +X_{d}(t-\tau) \Big\{ X_{s,0} -P_{s,0}\Big\} \notag  \\
& \hspace{1.6cm} +P_{d}(t-\tau) \Big\{ P_{s,0} +X_{s,0}\Big\} \Big],\notag
\end{align}
where the $\sigma_-$-polarized light field quadrature $X_{c}(t)$
is defined in analogy to \eeqref{eq:XPdfield}.

The Hamiltonian $H_{0,{\rm lin}}$ in \eeqref{eq:Hlin} includes the free evolution of the mechanical and atomic system as well as the optical lattice potential induced  by the driving laser on the $|s\rangle$-state
\begin{align}
H_{0,{\rm lin}}
=&\hbar \om_\rmm a_\rmm^\dag a_\rmm +\hbar \left( \omega_{\rm at}+ \frac{\Omega_{\rm OL} }{2} \right) \sum_j \sigma^j_{\rm ss}. 
\label{eq:freeHamiltonian}
\end{align}

From \eeqref{eq:Hlin} we see that only the $\sigma_+$-field can mediate the interaction between the two systems. The $\sigma_-$-field associated to the driving laser just couples  to the mechanical oscillator and can therefore not mediate interactions. 
However, after eliminating the light field the coupling of the mechanical oscillator to the  $\sigma_-$-field will result in a mechanical diffusion rate, as we will show in \secref{sec:decoherence}.
Moreover, spontaneous photon scattering from the $\sigma_-$-field will lead to atomic diffusion (in a three dimensional picture), see also \secref{sec:decoherence}.

\subsection{Phase Shift of Quantum Field} 
\label{sec:positionphaseshift}

The interaction part of the linearized Hamiltonian in \eeqref{eq:Hlin} can be rewritten as
\begin{align}
H_{\rm int}^{\rm lin}
 &=\hbar \sqrt{2} g_\rmm \left[ X_{c}(t)+\frac{1}{\sqrt{2}}\left\{ \overline{X}_{d}(t)- \overline{P}_{d}(t)\right\}\right] X_\rmm \notag \\
 &+\hbar \sqrt{2 N} g_\rmat \Big[ \left\{  \overline{X}_d(t-\tau) - \overline{P}_d(t+\tau)\right\} X_{s,0}   \notag \\
  &\hspace{1.6cm}+\left\{  \overline{X}_d(t+\tau) + \overline{P}_d(t-\tau)\right\} P_{s,0} \Big]  ,
 \label{eq:HlinOLbasis} 
\end{align}
where we changed the basis for the $\sigma_+$-polarized quantum light field to
\begin{align}
\overline{X}_d(t)&= \left\{X_{d}(t) + P_{d}(t)\right\}/\sqrt{2} \\
\overline{P}_d(t)&=\left\{ -X_{d}(t)+ P_{d}(t)\right\}/\sqrt{2}
\end{align}
with the commutator $[X_d(t),P_d(t')]=[\overline{X}_d(t),\overline{P}_d(t')]=i \delta(t-t')$.\\ 

\begin{figure}[h!]
\centering
\includegraphics[width=0.5\textwidth]{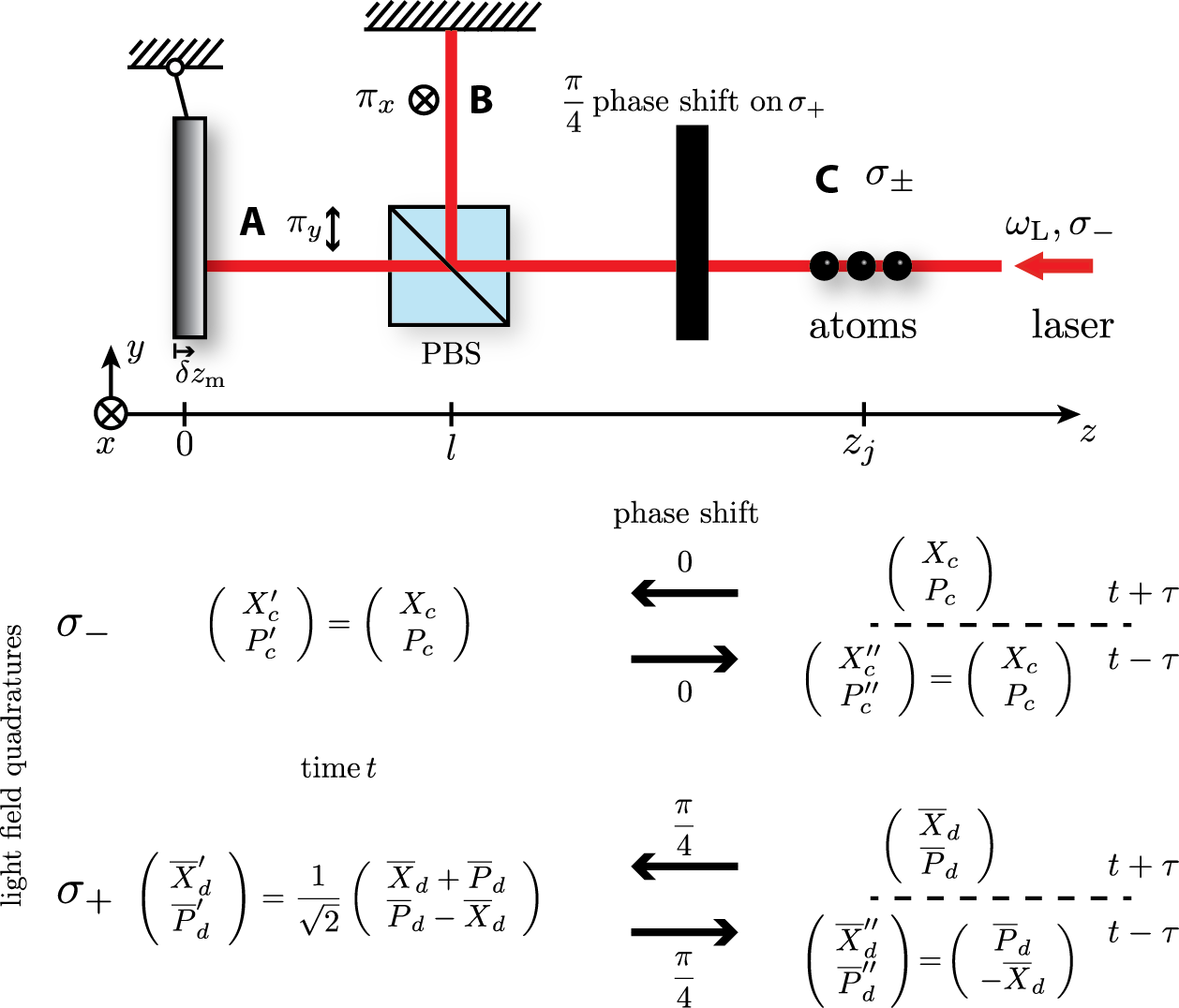}
\caption{Modified setup with a phase shift on the $\sigma_+$-light. Below: change of quadrature basis due to the phase shift on $\sigma_+$-light field at different times $t\pm\tau$ and $t$.}
\label{fig:modsetup}
\end{figure}

From \eeqref{eq:HlinOLbasis} we can derive an effective interaction between mechanical resonator and the $S_0$ spin mode of the atomic ensemble. 
However, the resulting Hamiltonian contains in addition to the coupling between the ensemble and the mechanical resonator an undesirable contribution resulting from the backaction of the atomic ensemble on itself, cf. \appref{app:elimination}. 
Since we are interested in a coherent interaction between ensemble and mechanical oscillator, we have to modify the setup slightly in order to remove this backaction term.

In the following, we describe in detail a method to get rid of the back action term. 
From \eeqref{eq:HlinOLbasis} one can clearly see, that the atoms (here for $X_{s,0}$) couple at different times $t\pm \tau$ to different quadratures of the light field. 
However, to obtain a resonant interaction between atoms and mechanics mediated by the light field, we need a slightly different coupling. 
In particular, the cascaded interaction we consider here has three time steps, i.e. $t+\tau$, $t$ and $t-\tau$. 
At each time step a different subsystem interacts with the light field by coupling to one of its quadratures. 
In our case we have the following sequence of subsystems: atoms, mechanical oscillator, atoms. 
We aim to achieve a situation, where in the first time step the atoms couple to (for instance) $\overline{P}_d(t+\tau)$. 
Then, in the next time step, the mechanical oscillator has to couple to the canonical conjugated quadrature $\overline{X}_d(t)$ to "read out" the information of the previous interaction. 
Finally in the last time step, the atoms again have to couple to the $\overline{P}_d(t-\tau)$-quadrature of the light field to obtain the informations about the interaction in the previous time step. 
From this argument it is also clear, that the atoms can not exchange information via the light field with themselves, since the both subsystems (i.e. atoms at times $t\pm \tau$) couple to the same light field quadrature.

To obtain the kind of interaction described above, we have to introduce a phase shift on the light quadratures of the $\sigma_+$-polarized light. In particular, we need a clockwise rotation of $\pi/4$.
Physically, this corresponds to a time retardation of $\pi/4$ of the standing  $\sigma_+$-polarized wave on arm $C$, cf. \figref{fig:modsetup}. Thus, when the amplitude of the $\sigma_-$-polarized standing wave is maximal, the amplitude of the $\sigma_+$-polarized standing wave vanishes. 

As displayed in \figref{fig:modsetup} we therefore introduce a phase shift on the light quadratures of the $\sigma_+$-polarized light. The classical $\sigma_-$-polarized light field is not affected by the phase shift and hence the basis stays the same for all time. Experimentally one could realize this phase shift by using a Faraday rotator, which puts both, a spatial and a time retardation onto the  $\sigma_+$-polarized field. However,
the corresponding spatial shift of the $\sigma_+$-field can be absorbed into the definition of the collective spin wave modes, and is not discussed here further.

In order to incorporate the time retardation phase shift into our formalism, we have to consider that at the different positions of the subsystems in our setup the basis of the quantum light field got rotated and therefore we couple to different bases in the Hamiltonian. Interactions with spatially separated subsystems can directly be translated into retardations in the Hamiltonian, which is a feature of the $1D$-treatment. Hence, we have that different times in the Hamiltonians correspond to different subsystems and thus a different basis. 

In particular we assume the following, cf. \figref{fig:modsetup}.
First, we start from the atomic ensemble at time $t+\tau$, where the Hamiltonian is not altered so far.
At the position of the mirror we  passed once through the Faraday rotator, and thereby the quadratures of the $\sigma_+$-field got rotated by $\pi/4$, such that we couple at time $t$ at the mirror to the new quadratures $\overline{X}_d'$ and $\overline{P}_d'$ as defined in \figref{fig:modsetup}. 
In the second step, at time $t-\tau$ another $\pi/4$-rotation is applied to the quantum field and therefore the interaction between light field and atomic ensemble couples to a third quadrature basis $\overline{X}_d''$ and $\overline{P}_d''$, cf. \figref{fig:modsetup}. 
The complete linearized Hamiltonian in an interaction picture w.r.t. $H_{\rm field}$ associated with the modified setup is then given by the modified interaction and the free evolution $H_{0, {\rm lin}}$ as defined in \eeqref{eq:freeHamiltonian}
\begin{align}
H_{\rm mod}
 =&H_{0, {\rm lin}}+\hbar \sqrt{2} g_\rmm \left[ X_{c}(t) +\overline{X}_{d}(t) \right]X_\rmm    \label{eq:Hmod}  \\
&+\hbar \sqrt{2 N} g_\rmat \left[ -\overline{P}_d(t+\tau) X_{s,0} + \overline{X}_d(t+\tau) P_{s,0} \right. \notag \\
&\hspace{1.9cm}\left. +\overline{P}_d(t-\tau) X_{s,0} - \overline{X}_d(t-\tau) P_{s,0}\right] \notag
\end{align}
Here, we expressed the primed bases ($\overline{X}_d'$,$\overline{P}_d'$) and ($\overline{X}_d''$,$\overline{P}_d''$) in the original basis ($\overline{X}_d$,$\overline{P}_d$) by using the transformations displayed in \figref{fig:modsetup}.
As we will show in the following section, after eliminating the light field $H_{\rm mod}$ provides an effective interaction between the mechanical mode and the collective spin excitation without a backaction of the atoms on themselves.

\subsection{ Effective Dynamics} 
\label{sec:effdyn}

We are interested in deriving the effective dynamics of the coupling between the micro-mechanical resonator and the collective excitation of the atomic ensemble. 
Therefore, we make an adiabatic elimination of the field modes in a Born-Markov approximation that accounts for the cascaded character of the system and is done in the framework of quantum stochastic Schr\"odinger equation by using similar methods as in Ref. \cite{Hammerer2010}. The main results are presented in the following section and details of the calculation can be found in \appref{app:adiabaticelimination}. 

\subsubsection{Effective Master equation}

Whereas the detailed calculations can be found in \appref{app:adiabaticelimination}, we provide the resulting effective master equation:
\begin{align}
\dot\rho = -\frac{i}{\hbar} [H_{\rm eff},\rho] &+ \gamma_\rmm^{\rm diff} \mD\left[X_\rmm\right]\rho,
\label{eq:MEmod}
\end{align}
where the Lindblad contribution is defined by $\mD[x]\rho=x\rho x^\dag - \frac{1}{2}\{x^\dag x,\rho\}$ and the mechanical diffusion rate by 
\begin{align}
\gamma_\rmm^{\rm diff}=2 g_\rmm^2.
\label{eq:gammamdiff}
\end{align} 
The effective Hamiltonian reads
\begin{align}
H_{\rm eff}=&\hbar \om_\rmm a_\rmm^\dag a_\rmm+\hbar \left( \omega_{\rm at}+\frac{\Omega_{\rm OL}}{2}\right) \sum_j \sigma^j_{\rm ss}  \notag \\
& - \hbar  g_{\rm eff} X_\rmm X_{s,0}.
\label{eq:Heffcompletemod}
\end{align}
with effective coupling rate
\begin{align}
g_{\rm eff}=  2 \sqrt{N} g_{\rm at}g_\rmm= \sqrt{\frac{ N}{2}}  \frac{\Om_+ \Om_-}{ \Delta}   k_L \ell_\rmm.
\label{eq:geff}
\end{align}
where we introduced the Rabi frequencies $\Omega_\pm=\alpha \mE_{\omega_L} \mu_\pm/\hbar$.

\subsubsection{Decoherence}
\label{sec:decoherence}
In the previous section we concluded with a master equation for the effectively coupled mechanical oscillator-atomic ensemble system. As a result of the adiabatic elimination of the light field we already obtained the light-induced diffusion of the mechanical resonator $\gamma_\rmm^{\rm diff}$ due to the coupling to the field in form of the Lindblad term in \eeqref{eq:MEmod}.

The light-induced diffusion of the atomic ensemble drops out due to the $1D$-treatment as well as the phase shifts that we introduced in \secref{sec:positionphaseshift}, and therefore we have to add the proper diffusion of the atomic ensemble.

The atomic decoherence rate $\gamma_\text{at}^\text{diff}$ is the decoherence rate of a single collective excitation in the ensemble, i.e.\ if one atom is in the $|s\rangle$ state. This is equivalent to the single-atom photon scattering rate 
\begin{align}
\gamma_\text{at}^\text{diff}=\Gamma\frac{\Omega^2}{\Gamma^2+4\Delta^2+2\Omega^2},
\end{align}
 where $\Gamma$ is the spontaneous emission rate and $\Omega$ is the Rabi frequency of the strong $\sigma_-$-drive which can induce off-resonant scattering on transitions between $|s\rangle$ and the excited state $|e\rangle$, while $|g\rangle$ is a dark state. 

In addition to the light-induced diffusion, the system also faces thermal decoherence due to the mechanical oscillator coupling to its support, which is given by the coupling to a thermal bath at finite temperature $T_0$:
 \begin{align}
\dot\rho =& -\frac{i}{\hbar} [H_{\rm eff},\rho] +\gamma_{\rm m}^{\rm diff} D[X_{\rm m}] \rho +\gamma_{\rm at}^{\rm diff} D[S_{s,0}] \rho \notag \\
&+ \gamma_\rmm (N_\rmm+1) \mD[a_\rmm]\rho + \gamma_\rmm N_\rmm \mD[a_\rmm^\dag]\rho, 
\label{eq:MEAEthermal}
\end{align} 
  where the thermal decoherence rate is given by $ \gamma_\rmm N_\rmm\approx \frac{k_{\rm B} (T_0+T_{\rm eff})}{\hbar Q_\rmm} $ with Boltzmann constant $k_{\rm B}$, mechanical quality factor  $Q_\rmm=\omega_\rmm/\gamma_\rmm$, and effective temperature $T_{\rm eff}$ of the mechanical system due to laser heating \cite{Vogell2013}.

\section{Discussion and Applications}
\label{sec:applications}

In the previous sections, we reduced the description of our cascaded quantum system to an effective master equation in \eeqref{eq:MEAEthermal} that describes the coherent coupling between mechanical resonator and atomic ensemble. However, as discussed in \secref{sec:decoherence} these coherent dynamics are accompanied by several noise sources, such as thermal or light-induced diffusion. Our goal is to investigate regimes that yield interesting applications. In the following, we first discuss the possibility to extend the calculations to various optomechanical systems. Further, we present estimates on the coherent dynamics as well as sympathetic cooling. Then, we compare the coupling to internal states and the coupling to the motional atomic states, cf. \cite{Vogell2013}, and finally we present experimental parameters and realizations we have in mind.

\subsection{Generic Optomechanical System}
\label{sec:genericOMsystem}

As we have already shown in previous work \cite{Vogell2013}, the scaling with the mechanical-light coupling $g_\rmm$ is rather generic and also applies to optomechanical configurations with e.g high finesse cavities. 
In fact we can apply the presented theoretical model to any single-sided optomechanical system by choosing the corresponding mechanical-light coupling $g_\rmm$ from which one can infer the mechanical diffusion rate $\gamma_\rmm^{\rm diff}$ in \eeqref{eq:gammamdiff} and the effective coupling rate $g_{\rm eff}$ in \eeqref{eq:geff}.

The most general mechanical-light coupling rate in \eeqref{eq:gm} for a single-sided optomechanical cavity system is given by 
\begin{align}
g_\rmm^{\rm om}= \frac{2\alpha}{\sqrt{\pi}} \frac{g_0}{\kappa},
\end{align}
where $g_0$ is the general optomechanical single-photon coupling strength and $\kappa$ the cavity line-width. With this one could for instance calculate the mechanical-light coupling for a photonic-crystal optomechanical cavity ("zipper cavity") \cite{Cohen13,Eichenfield09}, see \secref{sec:parameters}.

In the lines of our previous work in Ref. \cite{Vogell2013} another interesting optomechanical system is the extension of the above derivations for the ideal metallic mirror onto a "membrane-in-the-middle``-setup as visualized in \figref{fig:fig1}(b). The resulting membrane-light coupling is then enhanced by the finesse $\mF$ of the cavity
 \begin{align}
g_{\rmm}^{\rm mim}= \alpha  \frac{k_\rmL \ell_\rmm}{ \sqrt{2\pi}} 2  \sabs{\mathfrak{r}_\rmm} \frac{2 \mF}{\pi},
\label{eq:GeffFinesse}
\end{align}
where $\mathfrak{r}_\rmm$ is the reflectivity of the membrane.

\subsection{Coherent Dynamics}
\label{sec:coherentdynamics}

As a first application we consider the observation of coherent dynamics between the mechanical oscillator and the spin wave excitation in the atomic ensemble. 
The coherent dynamics are induced by the interaction term \eeqref{eq:Heffcompletemod}.
In particular, we are interested in a regime, where the splitting of the two atomic ground states is resonant with the mechanical frequency, i.e. $\omega_{\rm at}+\Omega_{\rm OL}/2\approx \omega_{\rm m}$ as well as $\om_\rmm  \gg g_{\rm eff}$. In that case we can apply a rotating wave approximation to \eeqref{eq:Heffcompletemod} such that we obtain a beamsplitter-type interaction Hamiltonian 
\begin{align}
H_{\rm eff}&\approx - \hbar g_{\rm eff} \left(  a^\dag_\rmm S_{s,0} +S_{s,0}^\dag a_\rmm\right).
\label{eq:Hbeamsplitter}
\end{align}
This interaction allows for coherent transfer of single excitations between the spin wave and the mechanical mode. However, as discussed in \secref{sec:decoherence} the system suffers from various dissipation and diffusion processes. 
Hence for coherent transfer of single excitations it is required that the coupling rate exceeds all decoherence rates, which is expressed by the strong coupling conditions
\begin{align}
g_{\rm eff} \gg \gamma_{\rm at}^{\rm diff}, \gamma_{\rm m}^{\rm diff}, \gamma_{\rm m}^{\rm th},
\label{eq:strongcoupling}
\end{align}
where $\gamma_{\rm m}^{\rm th}=\gamma_\rmm N_\rmm$. Further, we define the total mechanical decoherence rate $\gamma_{\rm m}^{\rm tot}=\gamma_{\rm m}^{\rm th}+\gamma_{\rm m}^{\rm diff}$.

\begin{table}[htbp]
\begin{center}
\begin{tabular}{|c|c|}
\hline
$\om_\rmm$ & $2 \pi \cdot 10$ MHz  \\
\hline
$M$& $4 \cdot 10^{-14}$ kg\\
\hline
$Q_\rmm$&$10^5$\\
\hline
$g_0$&$ 2 \pi \cdot 1.83 $ MHz\\
\hline
$T_0$&$4$ K\\
\hline
$\Delta T_0/P$&$12$ K/mW\\
\hline
$\kappa$&$2 \pi \cdot 4$ GHz\\
\hline
\end{tabular}
\begin{tabular}{|c|c|}
\hline
$\om_\rmL$ & $2 \pi \cdot 378$ THz  \\
\hline
$\Delta$& $2 \pi \cdot 15$ MHz\\
\hline
$N$&$8.5 \cdot10^6$\\
\hline
$w_0$&$ 30 $ $\mu$m\\
\hline
$\mu$&$2.54 \cdot 10^{-29}$ C m\\
\hline
$\Gamma$&$2 \pi \cdot 5.75$ MHz\\
\hline
$P$ & $2.5 \cdot 10^{-7}$ W  \\
\hline
\end{tabular}
\end{center}
\caption{Set of optimized parameters based on the considerations in \secref{sec:applications} and \secref{sec:parameters}. The left column describes the mechanical parameters for the optomechanical system of a zipper-cavity \cite{Cohen13,Eichenfield09}, where $\Delta T_0/P$ describes the absorption heating due to the laser. The column on the right displays the atomic and laser parameters, where $\mu$ is the transition dipole matrix element of the $D1$-line of $^{87}{\rm Rb}$. The values for the detuning $\Delta$, the laser power $P$ and the beam waist $w_0$ are optimized within the given boundary conditions, and further details are given in \secref{sec:parameters}.}
\label{tab:parameter}
\end{table}

By examining \eeqref{eq:geff} we find that for a given optomechanical system coupled to the atomic ensemble
the coupling rate can be optimized by varying the detuning, the laser power or the beam waist of the laser. However, all of these parameters also alter the decoherence rates $\gamma_{\rm m}^{\rm tot}$ and $\gamma_{\rm at}^{\rm diff}$ significantly. 
Thus, identifying optimal values for beam waist, detuning and laser power results in a tradeoff between optimization of the strong coupling conditions on one side, and fulfilling the conditions for adiabatically eliminating the excited state $|e\rangle$ as well as keeping the Stark shift of the $|s\rangle$ as low as possible. 

\begin{figure}
\centering
\includegraphics[width=0.5\textwidth]{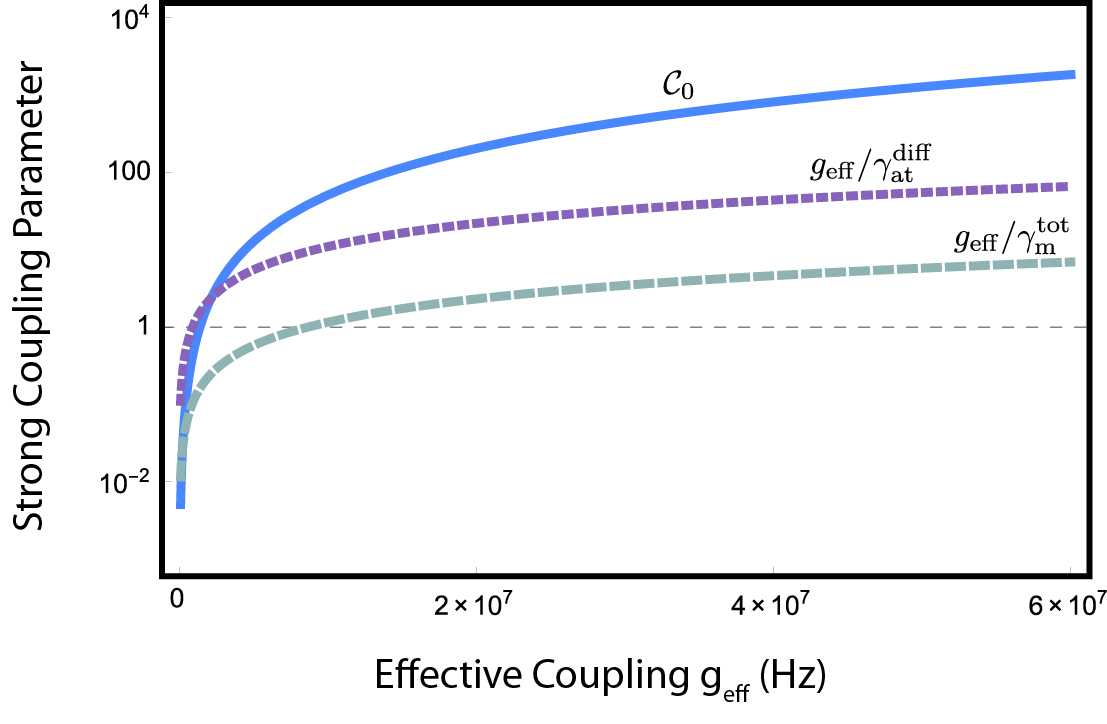}
\caption{Strong coupling conditions: we plot the ratios $g_{\rm eff}/\gamma_{\rm m}^{\rm tot}$ and $g_{\rm eff}/\gamma_{\rm at}^{\rm diff}$ as functions of the effective coupling $g_{\rm eff}$ as a benchmark for the strong coupling conditions for both, the mechanical and atomic decoherence. Further, the cooperativity $\mathcal{C}_0$ for coherent dynamics is displayed as a function of the effective coupling. Parameters are taken from Table \ref{tab:parameter}. 
}
\label{fig:strongcoupling}
\end{figure}

In order to show the fulfilment of the strong coupling conditions and therefore the ability to observe coherent dynamics, we display in \figref{fig:strongcoupling} the ratios $g_{\rm eff}/\gamma_{\rm m}^{\rm tot}$, as well as $g_{\rm eff}/\gamma_{\rm at}^{\rm diff}$
as functions of the effective coupling, where all other parameters are fixed by the optimized values in Table \ref{tab:parameter}.

We can clearly see that the larger the effective coupling the better the fulfilment of the strong coupling conditions. The upper bound for increasing the effective coupling is given by the rotating wave approximation that was made to obtain the Hamiltonian in \eeqref{eq:Hbeamsplitter}, which is only valid in the limit where $g_{\rm eff} \ll \omega_{\rm m}$. 
Note, since we fixed the values for beam waist, laser power and detuning to the parameters in Table \ref{tab:parameter}, increasing the effective coupling $g_{\rm eff}$ in \figref{fig:strongcoupling} corresponds in principle to varying the density of the atomic ensemble. From \figref{fig:strongcoupling} we therefore conclude that the atomic density should be as large as possible with the limit that the rotating wave approximation is still valid. In Table \ref{tab:parameter} we assume a reasonable high atomic density.

As a second figure of merit for strong coupling we define the cooperativity of the system as
\begin{align}
\mathcal{C}_0=\frac{4 g_{\rm eff}^2}{\gamma_\rmm^{\rm tot}  \gamma_\rmat^{\rm diff}},
\label{eq:Coop0}
\end{align}
and plot it as a function of the effective coupling in \figref{fig:strongcoupling}. We observe that for a wide range of values of the effective coupling $g_{\rm eff}$ the cooperativity is much larger than one for fixed values of beam waist, laser power and detuning. 
Interestingly, there is a range where  $\mathcal{C}_0 > 1$ but the strong coupling condition $g_{\rm eff}/\gamma_\rmm^{\rm tot} < 1$.
The cooperativity as defined in \eeqref{eq:Coop0} is only a figure of merit for the coherent dynamics, where for the case of cooling the mechanical oscillator we have to define a modified cooperativity as we will discuss in the following section. 

Concluding we find that the strong coupling conditions in \eeqref{eq:strongcoupling} are fulfilled in a wide range of parameters. In Table \ref{tab:couplingrates} we summarize the coupling and decoherence rates as well as the cooperativity for a set of optimized parameters given in Table \ref{tab:parameter}. These resulting strong coupling conditions are an improvement considering the coupling to the atomic motion, where the effective coupling and the decoherence rates were on the same order \cite{Vogell2013}.

\begin{table}[htbp]
\begin{center}
\begin{tabular}{|c|c|}
\hline
$g _{\rm eff}$ & $2 \pi \cdot 2.5$ MHz  \\
\hline
$\mathcal{C}_0$&$124.4$ \\
\hline
\end{tabular}
\begin{tabular}{|c|c|}
\hline
$\gamma_\rmm^{\rm diff}$ & $2 \pi \cdot 541$ kHz  \\
\hline
$\gamma_\rmat^{\rm diff}$& $2 \pi \cdot 143$ kHz\\
\hline
$\gamma_\rmm^{\rm th}$&$2 \pi \cdot 844$ kHz\\
\hline
\end{tabular}
\end{center}
\caption{Resulting coupling and decoherence rates as well as cooperativity after choosing optimal values for beam waist, laser power and detuning.}
\label{tab:couplingrates}
\end{table}

\subsection{Sympathetic Cooling}
\label{sec:sympatheticcooling}

The sophisticated atomic toolbox allows among other features to engineer dissipation. In particular we can prepare the atomic ensemble near to the ground state by repumping its population. Together with coherent interactions between atomic ensemble and membrane we obtain a sympathetic cooling effect on the membrane as was recently shown for coupling to the motional atomic degrees of freedom \cite{Jockel:2014kx}.

\begin{figure}
\centering
\includegraphics[width=0.45\textwidth]{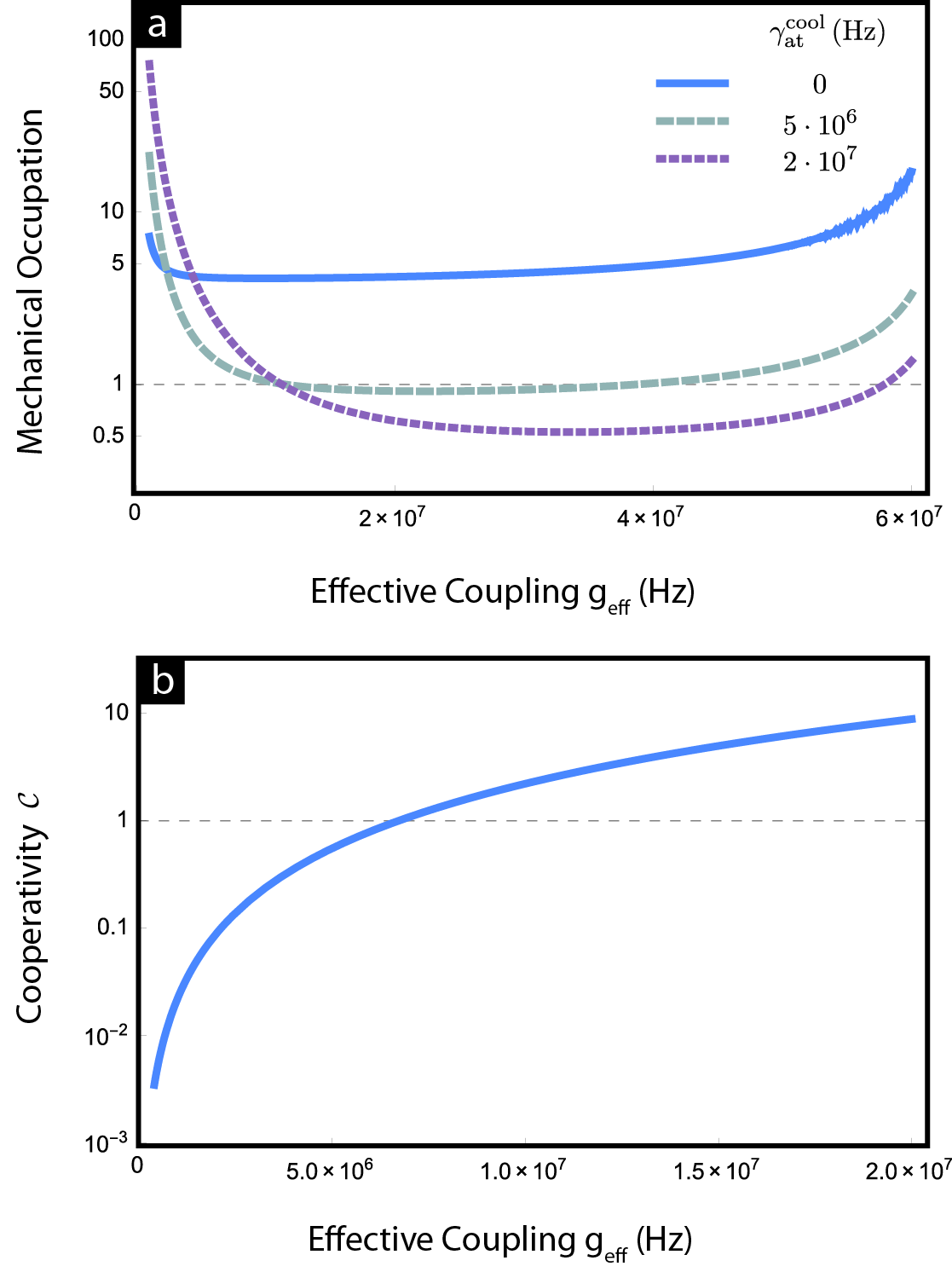}
\caption{Sympathetic cooling of the mechanical oscillator by externally re-pumping the atomic ensemble with rate $\gamma_{\rm at}^{\rm cool}$. 
(a) Steady state occupation of the mechanical oscillator as a function of the effective coupling is shown. The different curves correspond to different external re-pumping rates. 
Parameters are taken from Table \ref{tab:parameter}. 
(b) Cooperativity for sympathetic cooling setup as a function of the effective coupling is well above one for the optimized coupling in Table \ref{tab:couplingrates}. Here, an external cooling rate of $\gamma_{\rm at}^{\rm cool}=2 \cdot 10^7$ Hz is chosen. }
\label{fig:cooling}
\end{figure}

In \secref{sec:decoherence} we discussed the decoherence processes of the system. In particular we introduced the atomic dissipation due to the strong coupling laser, which results already in a cooling or better re-pumping process of the ensemble, cf. \eeqref{eq:MEAEthermal}. As a solid blue curve in \figref{fig:cooling}(a) we display the mechanical steady state occupation resulting from the solution of the master equation in \eeqref{eq:MEAEthermal}. Here, only the intrinsic re-pumping rate $\gamma_{\rm at}^{\rm diff}$  accounts for pumping the atomic ensemble to ground state $|g\rangle$. Thus we observe that for no external re-pumping laser the mechanical steady state occupation is already cooled down below ten quanta of excitation. 

However, to obtain ground state cooling of the mechanical oscillator we need to introduce an additional re-pumping laser on the side of the atomic ensemble, that impinges perpendicular to the quantization axis of the setup onto the ensemble. 
Incorporating this additional re-pumping laser into the master equation in \eeqref{eq:MEAEthermal} is done by introducing an amplitude decay with rate $\gamma_\rmat^{\rm cool}$ on the side of the atoms
 \begin{align}
\dot\rho =& -\frac{i}{\hbar} [H_{\rm eff},\rho] +\gamma_{\rm m}^{\rm diff} D[X_{\rm m}] \rho +\left(\gamma_{\rm at}^{\rm diff} +\gamma_{\rm at}^{\rm cool}\right) D[S_{s,0}] \rho \notag \\
&\qquad + \gamma_\rmm (N_\rmm+1) \mD[a_\rmm]\rho + \gamma_\rmm N_\rmm \mD[a_\rmm^\dag]\rho.
\label{eq:MEAEcooling}
\end{align}

In order derive the mechanical occupation number for the sympathetic cooling scheme 
we have to solve the full master equation in \eeqref{eq:MEAEcooling} with the Hamiltonian in \eeqref{eq:Heffcompletemod}.
In doing so we find that the system (for small $\gamma_\rmm$) is only stable in the limit
\begin{align}
\left(\gamma_{\rm at}^{\rm diff} +\gamma_{\rm at}^{\rm cool}\right)^2+4 \om_\rmm^2 > 4 g_{\rm eff}^2,
\label{eq:stability}
\end{align}
which provides a cut off for higher effective coupling rates in \figref{fig:cooling}(a). 
Whereas in the case of strong coupling as discussed in the previous section this inequality is automatically fulfilled, we have to include it when deriving the mechanical occupation. In principal this results in a cutoff for the effective coupling.

In \figref{fig:cooling}(a) we present the steady state occupation of the mechanical oscillator as a function of the effective coupling for three different external re-pumping rates $\gamma_{\rm at}^{\rm cool}$. Again we  clearly see that without external re-pumping (blue, solid) the mechanical oscillator can not be cooled to the ground state. 
In the previous section we introduced the optimized parameters in Table \ref{tab:parameter} for which the effective coupling takes the value $g_{\rm eff}=2 \pi \cdot 2.5$ MHz. Using this, we find that a re-pumping rate $\gamma_{\rm at}^{\rm cool}=2 \cdot 10^7$ Hz (purple, dotted) would allow for ground state cooling of the mechanical oscillator. 
Choosing a smaller re-pumping rate of $\gamma_{\rm at}^{\rm cool}=5 \cdot 10^6$ Hz (green, dashed) results in a mechanical steady state occupation exactly at the edge of the ground state. 

Further, by adding a re-pumping laser on the side of the atoms we have a modified atomic dissipation and therefore we have to redefine the cooperativity in \eeqref{eq:Coop0}. For the sympathetic cooling setup we define the cooperativity as 
\begin{align}
\mathcal{C}=\frac{4 g_{\rm eff}^2}{\gamma_\rmm^{\rm tot}  \gamma_\rmat^{\rm tot}},
\label{eq:Coop}
\end{align}
with the total atomic decoherence rate $\gamma_\rmat^{\rm tot}=\gamma_\rmat^{\rm diff}+\gamma_\rmat^{\rm cool}$. In \figref{fig:cooling}(b) we show the cooperativity for sympathetic cooling as defined in \eeqref{eq:Coop} as a function of the effective coupling rate. For small values of $g_{\rm eff}$ we clearly find that ground state cooling is not possible, which agrees with the numerical derivation of the mechanical steady state occupation in \figref{fig:cooling}(a).
Hence, we finally conclude that ground state cooling is possible for reasonable parameter regimes.

\subsection{Internal vs. motional States}
\label{sec:comparision}

Previously, systems that coupled the motion of the mechanical oscillator to the center-of-mass motion of the atomic ensemble \cite{Hammerer2010, Camerer2011, Vogell2013} have been investigated. In this manuscript we discussed a possible realization of coupling the internal degrees of freedom to the motion of the mechanical oscillator. In the following we discuss the important differences between the two different coupling schemes.

The coupling to the motional state of a harmonic oscillator leads to a Lamb-Dicke factor $\eta=k_L \ell_{\rm ZPF}\ll 1$ in the coupling. Since this factor is in general small, we profit in the case of internal state coupling from not having a Lamb-Dicke factor on the side of the atoms. We can demonstrate this by comparing the effective coupling rate for the internal states in the case of a "membrane-in-the-middle`` setup, \eeqref{eq:GeffFinesse} to 
the coupling of the center-of-mass mode to the mechanical oscillator from Eq. (27) in Ref.\cite{Vogell2013}. The latter is given by
\begin{align}
g_{\rm com}\propto \omega_{\rm at} \frac{\ell_\rmm}{\ell_{\rm at}} \abs{\mathfrak{r}_\rmm} \sqrt{N} \mF,
\end{align}
where $\ell_{\rm at}=\sqrt{\hbar/2 m \omega_{\rm at}}$ are the atomic zero point fluctuations. Calculating the quotient of both coupling rates yields
 \begin{align}
\frac{g_{\rm eff}}{g_{\rm com}} &\propto \frac{\Om_+ \Om_-}{ \Delta}   k_L \frac{\ell_{\rm at}}{\omega_{\rm at}} =\frac{1}{k_\rmL \ell_{\rm at}}=\frac{1}{\eta_{\rm at}},
\end{align}
where we used the definition of the optical lattice in Ref. \cite{Vogell2013}, i.e. $V_0 k_\rmL^2 = m_{\rm at} \omega_{\rm at}^2/2$ with lattice depth $V_0=\hbar \Om_+ \Om_-/ \Delta $.

In addition, note that the atomic-light diffusion $\gamma_\rmat^{\rm diff}$ in the motional coupling is proportional to the atom-field coupling $g_\rmat^2$ and thereby suppressed by a Lamb-Dicke factor squared. In the case of internal state coupling we loose this suppression of the diffusion rate by the Lamb-Dicke factor, and thereby we have a much higher atomic diffusion rate. Nevertheless this is not necessarily a drawback since the $\gamma_{\rm at}^{\rm diff}$ in \eeqref{eq:MEAEthermal} corresponds to a cooling rate rather than a diffusion process as in the motional coupling, cf. \cite{Vogell2013}.

Further, from the experimental point of view coupling to the internal atomic states has certain advantages compared to the motional coupling.
In the motional coupling, there is a resonance condition between the frequencies of mechanical mode and atomic center-of-mass mode. Optical lattice potentials are limited to trap frequencies of several hundred kHz. This restriction no longer holds for internal states, since the resonance condition depends on the splitting of the two ground states $|s\rangle$ and $|g\rangle$. In general,
the atomic levels offer a large range of energy splittings that could be addressed, e.g.\ Zeeman-sublevels split by magnetic fields in the MHz range or hyperfine ground states with splittings in the GHz range.
Thereby, the constraints on the mechanical frequency range are much more relaxed.
Finally, the internal state of the atoms can also be prepared and detected with a much higher fidelity than the center-of-mass motion.

\subsection{Experimental Parameters}
\label{sec:parameters}
In the following, we give more details on possible experimental realizations. On the atomic side, we consider a cloud of cold $^{87}$Rb atoms, which is routinely being prepared in numerous cold atom experiments. To calculate the number $N$ of atoms that effectively couple to the optomechanical system we assume a homogeneous density of $3 \cdot10^{17}/$m$^3$, which is one order of magnitude below the limit which can be reached with Raman sideband cooling \cite{Kerman2000}. A realistic length of the atomic cloud of 1\,cm yields an atomic area density of $\rho_\text{A}\approx 3 \cdot 10^{15}$/m$^2$. Assuming the beam has a circular shape with radius $w_0$, the atom number is given by $N=\rho_\text{A}\pi w_0^2$. 

We consider the mechanical mode to be coupled to Zeeman-split sublevels of a long-lived atomic hyperfine state. More specifically, we choose the two sublevels $|g\rangle=|F=2, m_F=-2\rangle$ and $|s\rangle=|F=2, m_F=0\rangle$ of the $5^2$S$_{1/2}$ ground state. Using an external magnetic field, the splitting between $|g\rangle$ and $|s\rangle$ can be tuned into resonance with the mechanical frequency $\omega_\rmm$. The levels are coupled by a weak $\sigma_+$-sideband and a strong $\sigma_-$-drive via the $|F'=1,2,  m'_F=-1\rangle$ sublevels of the $5^2$P$_{1/2}$ excited state (D$_1$ transition: $\omega_\text{L}/2\pi=378\,$THz, $\lambda=795\,$nm). If no sideband photons are present, the atoms will be optically pumped to the energetically lower $|g\rangle$-state. 

Concerning the optomechanical system, we first consider a photonic-crystal optomechanical cavity (``zipper cavity'') which was developed in the group of O.\ Painter \cite{Cohen13,Eichenfield09}. These fibre-coupled nano-structured devices combine low effective masses on the picogram scale with strong field gradients on the wavelength scale to obtain huge single-photon coupling strengths $g_0$. As realistic parameters for a zipper cavity we assume a mechanical mode with frequency $\omega_\rmm/2\pi=10\,$MHz and $g_0/2\pi=1.83\,$MHz, cavity linewidth $\kappa/2\pi=4\,$GHz, $M=4\times10^{-14}\,$kg, $Q_\rmm=10^5$, see Table~\ref{tab:parameter}. We further assume a $^4$He cryogenic environment with $T_0=4\,$K to minimize thermal dissipation. In order to model the heating due to the laser drive as discussed in \secref{sec:decoherence} we assume an absorption heating of $\Delta T_0/P= 12\, {\rm K/mW}$ (estimation based on measurements in \cite{Eichenfield09}).

As optimization parameters to observe coherent dynamics (see \secref{sec:coherentdynamics}), we vary the beam power $P$, the detuning $\Delta$ from the excited $F'=2$ state and the beam radius $w_0$. We find an effective coupling rate $g_\text{eff}=15.6\,$MHz which is about one order of magnitude higher than the atomic, mechanical and thermal diffusion rates $(\gamma_\rmat^\text{diff}, \gamma_\rmm^\text{diff}, \gamma_\rmm^\text{th})=(0.9,3.4,5.3)$\,MHz, cf. Table \ref{tab:couplingrates}. These optimum values are achieved by tuning the laser almost on resonance with the atoms ($\Delta/2\pi=15\,$MHz) and making the beam as small as possible ($w_0=30\,\mu$m), i.e.\, keeping the Rayleigh range longer than the atomic ensemble. The power $P=2.5\cdot10^{-7}$W is just below atomic saturation but far below the range where we expect significant heating of the zipper cavity.

Second, we consider a "membrane-in-the-middle`` setup similar to that described in \cite{Jockel:2014kx}. While the membrane properties are very similar to the ones in \cite{Jockel:2014kx}, we assume a more compact cavity with a length of about 1\,mm and a higher finesse of $\mathcal{F}=700$ which is placed in a cryogenic environment at $T_0=4\,$K, see Table~\ref{tab:parameter_mim}. Heating by absorption of laser light $\Delta T_0/P$ is modeled as described in \cite{Vogell2013}.

\begin{table}[htbp]
\begin{center}
\begin{tabular}{|c|c|}
\hline
$\om_\rmm$ & $2 \pi \cdot 276$\,kHz  \\
\hline
$M$& $4 \cdot 10^{-10}$\,kg\\
\hline
$Q_\rmm$&$1.9\cdot10^6$\\
\hline
$\mathfrak{r}_\rmm$ & $0.4$\\
\hline
$\mathcal{F}$ & $700$\\
\hline
$g_0$&$ 2 \pi \cdot 175$\,Hz\\
\hline
$T_0$&$4$\,K\\
\hline
$\Delta T_0/P$&$2.2$\,K/mW\\
\hline
$\kappa$&$2 \pi \cdot 232$\,MHz\\
\hline
\end{tabular}
\end{center}
\caption{Properties of the "membrane-in-the-middle`` system.
\label{tab:parameter_mim}}
\end{table}

Analogous to the zipper cavity, we find optimum parameters to observe coherent dynamics using the MIM setup, see Table~\ref{opt_par_mim}. Here, compared to the zipper cavity, we are less limited by the mechanical diffusion rates, which allows us to use a much higher laser power. However, in order to keep the atomic excitation small, we need a much larger detuning and/or a larger beam radius. Finally, the coherent coupling rate is one order of magnitude smaller than for the zipper cavity and it is only slightly larger than the decoherence rates, placing the system at the edge of strong coupling.

\begin{table}[htbp]
\begin{center}
\begin{tabular}{|c|c|}
\hline
$\Delta$& $2 \pi \cdot 1.1$\,GHz\\
\hline
$w_0$&$ 50\,\mu$m\\
\hline
$P$ & $2.5\,$mW  \\
\hline
$N$&$2.4\cdot10^7$\\
\hline
\end{tabular}
\begin{tabular}{|c|c|}
\hline
$g _{\rm eff}$ & $2 \pi \cdot 150$\,kHz  \\
\hline
$\mathcal{C}_0$&$6.5$ \\
\hline
\end{tabular}
\begin{tabular}{|c|c|}
\hline
$\gamma_\rmm^{\rm diff}$ & $2 \pi \cdot 15$\,kHz  \\
\hline
$\gamma_\rmat^{\rm diff}$& $2 \pi \cdot 113$\,kHz\\
\hline
$\gamma_\rmm^{\rm th}$&$2 \pi \cdot 105$\,kHz\\
\hline
\end{tabular}
\end{center}
\caption{Left column: Set of optimized parameters to observe coherent dynamics in the "membrane-in-the-middle`` setup. Middle column: Coherent coupling rate and cooperativity. Right column: Decoherence rates.}\label{opt_par_mim}
\end{table}

\section{Conclusion}
\label{sec:conclusion}

In summary, in this paper we have discussed the full quantum model for a hybrid quantum system consisting of a mechanical resonator coupled to the internal states of an atomic ensemble. 
Coupling in particular to the internal states of the atoms rather than the motional states offers many advantages like tunability of frequencies and full access to the atomic toolbox. 
We exploit these features of the internal state coupling and present, in addition to the coherent dynamics, that the pre-cooled mechanical oscillator can be cooled to its ground state by sympathetic cooling via the atomic ensemble. 
Further, the quantum model is not limited to a specific mechanical system, but can be generalized onto various mechanical resonators such as membranes or photonic crystal cavities. We conclude our derivations by comparing the proposal to previous work on motional state coupling \cite{Vogell2013}.

We thank K. Stannigel and A. J\"ockel for helpful discussions. 
We acknowledge support from the ERC-Synergy Grant UQUAM, the SFB FOQUS of the Austrian Science Fund, the European
Commission (iQUOEMS), the Swiss Nanoscience Institute, the NCCR Quantum Science and Technology and the European project SIQS. 
M.T.R. acknowledges support from a Marie Curie IIF fellowship.

\appendix
\section{Hamiltonian without Phase Shift}
\label{app:elimination}

From \eeqref{eq:HlinOLbasis} we can derive an effective interaction between mechanical resonator and the $S_0$ spin mode of the atomic ensemble by eliminating the light field.
In doing so, we eliminate the mediating light fields of $\sigma_\pm$-polarized light using the formalism described in \appref{app:adiabaticelimination}.
Finally, the resulting Hamiltonian then reads
\begin{align}
H_{\rm eff}^{\rm lin}=-\hbar \sqrt{2 N} g_\rmm g_\rmat X_\rmm X_{s,0}-\hbar N g_\rmat^2( X_{s,0}^2+P_{s,0}^2).
\label{eq:HwoPS}
\end{align}
In addition to the coupling of the ensemble to the mechanical resonator the effective atom-mechanics interaction Hamiltonian in \eeqref{eq:HwoPS} has a second contribution from the backaction of the atomic ensemble on itself. This atom-atom interaction is enhanced by the number of atoms and thereby much stronger than the atom-mechanics interaction term. Hence, for the purpose of coherent interaction between atomic ensemble and mechanical resonator, it is undesirable to have this contribution. Modifying the setup slightly by introducing phase shifts on the quadratures of the quantum field is one way to resolve this problem and remove the atom-atom backaction term.

\section{Adiabatic Elimination}
\label{app:adiabaticelimination}

In the following, we summarize the calculations leading to an effective description of the mechanical oscillator coupled to the atomic ensemble. 

Therefore, we start with the fully linearized Hamiltonian in \eeqref{eq:Hmod} 
that is in a rotating frame with respect to $H_{\rm field}$.
The Hamiltonian includes the interaction between the mechanical resonator as well as the atomic ensemble with the light field. 
The system is then governed by the Schoredinger equation
\begin{align}
\frac{d}{dt} |\Psi\rangle&=-\frac{i}{\hbar} H_{\rm mod}(t) |\Psi\rangle \label{eq:QSSE}.
\end{align}
 Note that all optical frequencies are removed from $H_{\rm mod}(t)$ and further that we assumed all photons mediating the relevant interactions have frequencies in a band width $2\theta$ around the laser frequency $\omega_{\rm L}$.

We are interested in a situation, where $\theta \rightarrow \infty$ such that the field operators in \eeqref{eq:commutationrel} become $\delta$-correlated. In this so-called white noise limit we can interpret \eeqref{eq:QSSE} as a quantum stochastic Schr\"odinger equation (QSSE) of Stratonovich type with time delays \cite{quantumnoise}. By integrating \eeqref{eq:QSSE} in small time steps $\Delta t$ up to second order in the interaction Hamiltonian, we obtain an effective interaction between atomic ensemble and mechanical resonator, that is mediated by the photons. 
This is basically a Born-Markov approximation in the coupling between subsystems (atomic/mechanical) and field. We further assume that the field is initially in the vacuum state. 
By taking the time ordering of the interactions into account and further taking the limit $\Delta t \rightarrow 0$, we end up with a QSSE of Ito-type {\it without} time delays. This result can then be used to derive an effective master equation for the coupling between mechanical and atomic subsystem. 

Summarizing the complete hierarchy of timescales we find
\begin{align}
\frac{1}{\om_{\rm L}} \ll \frac{1}{\theta} \ll \abs{\tau} \ll \Delta t \ll \frac{1}{g_\rmm^2},\frac{1}{g_\rmat^2}, \frac{1}{\omega_{\rm m,at}},
\label{eq:hierarchy}
\end{align}
where we have $\om_{\rm L} \gg \theta \gg \om_{\rm m,at}$ from the definition of the bandwidth $\theta$ in \eqsref{eq:Defc}-\eqref{eq:Defd}. Further, the time interval $\Delta t$ of the integration is much longer than the propagation times $\tau$ between the systems, but much shorter than the system timescales. Finally, $1/\theta \ll \abs{\tau}$ is needed to be able to distinguish the temporal order of photon emission and reabsorption events in the interaction.

\subsection{Solving QSSE (Stratonovich type) with time delays} 

From the Schr\"odinger equation in \eeqref{eq:QSSE} we obtain the time evolution of the state $|\Psi(t_0)\rangle$ for some initial time $t_0$ by
\begin{align}
|\Psi(t_0+\Delta t)\rangle= U(t_0, \Delta t) |\Psi(t_0)\rangle
\label{eq:timeevolution}
\end{align}
with time evolution operator for the time internval $\Delta t$
\begin{align}
U(t_0, \Delta t)= e^{-\frac{i}{\hbar} \int_{t_0}^{t_0+\Delta t}\! \textnormal{d}s \, H_{\rm mod}(s)}.
\end{align}
We assume that the time interval $\Delta t$ is much longer than the time delays $\tau_i$ and much shorter than the system evolution, i.e. coupling strength and decays, cf. \eeqref{eq:hierarchy}.
We expand the right hand side of \eeqref{eq:timeevolution} in the time interval $\Delta t$ up to second order in the Hamiltonian:
\begin{align}
&U(t_0, \Delta t) |\Psi(t_0)\rangle \notag \\
&\approx  \left \{ 1- \frac{i}{\hbar}  \int_{t_0}^{t_0+\Delta t} \hspace{-0.75cm} \textnormal{d}s  \, H_{\rm mod}(s) \right. \label{eq:timeevo} \\
&\hspace{0.9cm}\left. -  \frac{1}{\hbar^2}  \int_{t_0}^{t_0+\Delta t} \hspace{-0.75cm} \textnormal{d}s \, H_{\rm mod}(s) \int_{t_0}^{s} \hspace{-0.2cm} \textnormal{d}s' \,  H_{\rm mod}(s')\right \} |\Psi(t_0)\rangle,\notag
\end{align}
where the bounds of the second integral in the third term account for the time ordering. We expand the above expression up to first order in $\Delta t$. Further, we assume that the field is initially in a vacuum state such that the initial wave function reads $|\Psi(t_0)\rangle=|\Psi(t_0)\rangle_{\rm sys}\otimes |0\rangle_{\rm field}$. As a result, we neglect all terms with annihilation operators acting directly on the initial state as well as terms with two consecutive creation operators, because they are of higher order \cite{Hammerer2010}. Evaluating the second term in \eeqref{eq:timeevo} yields:
\begin{align}
&-i \left \{ 
\frac{H_{0,{\rm lin}}}{\hbar}  \Delta t +
 g_{\rm m} X_\rmm \left[ \Delta C^\dag(t_0,0)+\Delta D^\dag(t_0,0) \right] \right.
\notag \\
&\hspace{0.5cm}
 +i \sqrt{N} g_{\rm at} X_{s,0} \left[ \Delta D^\dag(t_0,-\tau)- \Delta D^\dag(t_0,\tau) \right] \notag \\
&\hspace{0.5cm}\left.
 + \sqrt{N} g_{\rm at} P_{s,0} \left[ \Delta D^\dag(t_0,\tau)- \Delta D^\dag(t_0,-\tau) \right]
\right \}
 |\Psi(t_0)\rangle.
\end{align}
Here, we introduced the noise-increment operators
\begin{align}
 \Delta D^{\dag}(t_0,\pm \tau)&=\int_{t_0}^{t_0+\Delta t}\hspace{-0.5cm}\textnormal{d}s \, d^{\dag}(s\pm\tau) \\
 \Delta C^{\dag}(t_0,\pm \tau)&=\int_{t_0}^{t_0+\Delta t}\hspace{-0.5cm} \textnormal{d}s \, c^{\dag}(s\pm\tau)
  \end{align}
with commutation relations
\begin{align}
\left[ \Delta D(t_0,\tau_1),\Delta D^{\dag}(t_0,\tau_2) \right]= \Delta t- (\tau_1-\tau_2) \approx \Delta t,
\label{eq:commRelD}
 \end{align}
and same for $\Delta C(t)$. In the last step of \eeqref{eq:commRelD}, we assumed that the time delays $\tau_i$ of the system are much smaller than the considered time interval $\Delta t$. 

The third term in \eeqref{eq:timeevo} 
has four contributions from evaluating the product of the two Hamiltonians $H_{\rm mod}$. 
Since we only keep contributions up to first order in $\Delta t$, we can directly drop three of these terms, which leaves us with
\begin{align}
& -  \left[ g_\rmm^2 X^2_\rmm -i 2 \sqrt{N} g_{\rm at}g_\rmm X_\rmm X_{s,0} \right] \Delta t  |\Psi(t_0)\rangle,
\label{eq:2ndcontrib_timeevo}
\end{align}
where we used that the field is initially in the vacuum state, such that $d(t) d^\dag(t') |\Psi(t_0)\rangle=\left[d(t), d^\dag(t')\right] |\Psi(t_0)\rangle$.
 
 In the end we want to have a differential form of \eeqref{eq:timeevolution}, which is achieved by sending $\Delta t \rightarrow 0$. In order to do so, we have to maintain the hierarchy of timescales in \eeqref{eq:hierarchy}. Since the first two timescales ($1/\om_{\rm L}, 1/\theta$) already disappeared from the problem, we start by sending $\tau \rightarrow 0$. The latter is equivalent to neglecting the retardations, which leads to
$\Delta D^\dag(t_0,\pm \tau)=\Delta D^\dag(t_0,0)\equiv \Delta D^\dag(t_0)$. 
Finally, we take the limit $\Delta t \rightarrow 0$ such that we can rewrite \eeqref{eq:timeevolution} in differential form.
The result then gives us the time evolution for an interval $dt$. We assume that each time interval does not depend on the previous one (Markov approximation), and therefore the above result is valid for all $t$, such that we can write
\begin{align}
\label{eq:timeevo5}
d|\Psi\rangle&=|\Psi(t+dt)\rangle - |\Psi(t)\rangle  \\
&=-i \Big \{\left(\frac{H_{0,{\rm lin}}}{\hbar}-  i g_\rmm^2 X^2_\rmm - 2  \sqrt{N} g_{\rm at}g_\rmm X_\rmm X_{s,0} \right) dt \notag \\
&\hspace{1cm}+ \left[ g_{\rm m} X_\rmm \left[ dC^\dag(t)+dD^\dag(t) \right] \right]  \Big \} |\Psi(t)\rangle,\notag
\end{align}
where $dD(t)=\lim\limits_{{\Delta t \rightarrow 0}}{} \Delta D(t)$ and similar for $d C(t)$.
These noise-increment operators $dD(t)$ fulfill the Ito table for fields in vacuum, i.e. $dDdD=dD^\dag dD^\dag=0$ and $dD dD^\dag=dt$ \cite{quantumnoise}.

\subsection{Effective Master Equation} 

In order to obtain the master equation from \eeqref{eq:timeevo5}, we consider the evolution of the full density matrix and subsequently trace over the light field. 
According to the Ito calculus, we have $d\hat{\rho}=d\left(|\Psi\rangle\langle \Psi|\right)=(d|\Psi\rangle)\langle \Psi|+|\Psi\rangle (d\langle \Psi|)+(d|\Psi\rangle)(d\langle \Psi|)$, where $\hat{\rho}$ is the density matrix of the full system. We insert \eeqref{eq:timeevo5} in this expression and neglect all contributions of the order higher than $ dt$. Subsequently, we trace over the field degrees of freedom with
\begin{align}
&\rm{Tr}_{\rm fi}(dC^\dag \hat{\rho})=\rm{Tr}_{\rm fi}(dC \hat{\rho})=\rm{Tr}_{\rm fi}(dD^\dag \hat{\rho})=\rm{Tr}_{\rm fi}(dD \hat{\rho})=0 \notag \\
&\rm{Tr}_{\rm fi}(dC^\dag \hat{\rho} dC)=\rm{Tr}_{\rm fi}(dD^\dag \hat{\rho} dD) \approx \rho dt,
\end{align}
where we used the Ito table for vacuum field and defined $\rho=\rm{Tr}_{\rm fi}(\hat{\rho})$ as the reduced density matrix for the atom-mechanics system. This results in the master equation in \eeqref{eq:MEmod} with effective Hamiltonian in \eeqref{eq:Heffcompletemod}.


\bibliography{atommembranebibliography}

\end{document}